# Psychological Types of Brazilian Software Engineering Students

Luiz Fernando Capretz
University of Western Ontario
Dept. Electrical & Computer Engineering
London, Ontario, N6G 1H1, Canada
Tel: +1-519-6612111 (x85482)
lcapretz@eng.uwo.ca

**Abstract**

*The aim of this investigation was to establish the personality profile of Brazilian software engineering students according to the MBTI. This study also shows that the software engineering field attracts students of some types more than other types, for instance: Is, Ps, IPs, TPs, and INs are significantly represented in that group as opposed to E, Js, EJs, TJs, ENs.*

**Bio-sketch**

**Luiz Fernando Capretz** has over 20 years of experience in the software engineering field as a practitioner, manager and educator. Before joining the University of Western Ontario, in Canada, he worked at both technical and managerial levels, taught and carried out research on the engineering of software in Brazil, Argentina, England and Japan. He was the Director of Informatics and Coordinator of the Computer Science program in two universities (UMC and COC) in the State of Sao Paulo/Brazil. He has authored and co-authored over 50 peer-reviewed research papers on software engineering in leading international journals and conference proceedings, and co-authored the book, *Object-Oriented Software: Design an Maintenance*, published by World Scientific. His current research interests are software engineering (SE), human factors in SE, software quality, software product lines, and software engineering education. Dr. Capretz received his Ph.D. from the University of Newcastle upon Tyne (U.K.), Master's from the National Institute for Space Research (INPE - Brazil), and B.Sc. from UNICAMP (Brazil). He is a senior member of IEEE, and a MBTI Qualified Practitioner.

**Psychological Types of Brazilian Software Engineering Students**

The software industry has become a major force in today's society, but software engineering is a field that many outsiders and even insiders have strongly stereotyped. It is commonly believed that to be a good software developer it is necessary to like mathematics or a similar field. People stereotype the behaviour of software professionals, seeing them as introverts working alone in a corner of their office, hating interaction with others; in other words, as typical *nerds*. However, specialties within software engineering today are as diverse as the medical profession, with software engineers working as systems analysts, interface designers, programmers, testers, maintainers, help-desk trouble shooters, and so forth.

When people speak of software, they may be referring to the structure of a program, the functionality of an application system, the look and feel of an interface or the overall user experience with a hardware-software environment. Software engineering includes both new software developments and the maintenance of legacy systems; each software life cycle phase brings its own contexts of understanding about what matters, what can be designed, and what tools and methods are appropriate.

<u>System Analysis</u>: The system analysis phase emphasizes identification of high-level components in a real-world application and decomposition of the software system. The system analysis phase requires that the system analyst:

- understands the system's essential features;

- considers the requirements that need to be satisfied by the software system;
- creates an abstract model of the application in which these requirements are met.

It is during this phase that an abstract model of the application that comprises high-level abstractions of software components is best understood. The main product of the system analysis phase is a graphical or textual description (informal or formal) of an abstract model of the application. System analysis also requires a great deal of human interaction with users and clients.

<u>Design</u>: Design is an ambiguous term. Although there is huge diversity among design principles, common concerns and principles that are applicable to the design of any artefact can be found: whether it is a poster, a household appliance, or a housing development. In particular, software design is still a young field, and we are far from having a widely accepted definition of its relevant principles. Software design should be a user-oriented field, and as such will always include the human element possessed by other disciplines such as architecture or graphic design, rather than the hard edge formulaic certainty of engineering design.

Software design is an exploratory process. The designer looks for components by trying out a variety of schemes in order to discover the most natural and reasonable way to refine the application. Software design has been shown in such a manner that it appear simple to create. Nevertheless, in the design of large and complex software, identification of key components is likely to take some time. Repetitions are not unusual since a good design usually takes several iterations. The number of iterations also depends on the designer's insight and experience in the application domain.

Programming: Programming involves translating a refined version of the design into a programming language in order to provide the overall software services. This phase requires identification of control structures, relevant variables and data structures, and a detailed understanding of the syntax and specifics of a programming language. This usually follows an iterative stepwise refinement process that is mostly top-down, breadth first. Therefore, programmers need to attend to details and maintain an open, logical, analytical style.

Testing: This phase involves finding and correcting (debugging) errors in a program. The testing stage may not be the first time when errors appear; they can be carried through from the system analysis and design phases. However, testing is focused on finding faults, and there are many ways to make testing efforts more efficient and effective.

First, each module is tested on its own, isolated from the other components in the system. This testing, known as unit testing, verifies that a module functions properly with the various input expected (and unexpected!) based on the module's design. After collections of modules have been unit-tested, the next step is to ensure that the interfaces among them are well defined. Integration testing is the process of verifying whether the system components work properly together. Testing strategies are neither random nor haphazard but should be approached in a methodical and systematic manner. Once a fault is detected, debugging can be a frustrating and emotionally challenging activity that can lead software engineers to restructure their thinking and decisions. This requires persistence in choosing from an enormous range of possibilities and maintaining a higher level of attention to detail.

Maintenance: Software is normally subject to continuing changes after it is written and delivered. Thus, the challenge now becomes one of maintaining a continually evolving operational system.

Bishop-Clark (1995) investigated the relationship between cognitive aspects, personality traits and computer programming. She divides programming in several stages: problem representation, program design, implementation and debugging. She organized the theories and the empirical studies of computer programming into four sub-tasks: problem solving, designing, coding, and debugging. The cognitive styles discussed in some detail include: field dependency/independency, analytic/holistic, impulsivity/reflectivity, and divergent thinking; the personality traits include locus of control, and introversion/extroversion. These variables were mentioned because, according to her theory, they were all important within the realm of computer programming.

Cognitive styles have been studied as factors that may help explain some of the variability; however, they have failed to consistently explain individual preference towards computer programming as opposed to system analysis, for example. MBTI offers a potential to provide a suitable model for comparison.

Several empirical studies have investigated the relationship between the MBTI and programming, mostly in the United States. For example, Sitton and Chmelir (1984) listed some stereotypes about what programmers are like and what attracts them to the field. They painted a picture of creative professionals merrily and irreverently solving complicated problems, untrammelled by routine and humdrum details; however, they gave no specific statistics to support their findings. Bush and Schkade (1985) tested 58 professionals in one high-tech aerospace company involved with scientific programming only. They found that ISTJ (25%) was the most common type. Further, the second most frequently reported type was INTJ (16%), with ENTP (9%) third, thinking (74%) and judging (70%) were well represented. Buie (1988) examined a sample of 47 scientific programmers employed by a private company under contract with NASA, all performing work on orbit-related software. ISTJ (19%), INTP (15%), and INTJ (13%) were the most frequent types, with those three types collectively accounting for nearly half the sample. ESFJ (0%), ISFP (0%), and ENTP (0%) were particularly underrepresented.

Nevertheless, there is more to software engineering than programming. The engineering of software comprises systems analysis, design, programming, testing, and maintenance of software systems; each of which demands different abilities. Indeed, each phase involves varied tasks that require different skills. For instance, the skills and activities involved in designing a software system are quite different from the skills and activities necessary to test the software properly.

Lyons (1985) surveyed 1,229 computer professionals (such as: programmers, analysts, engineers, and managers) employed by over 100 different companies in the United States, Australia and Great Britain, including insurance companies, financial institutions, utilities, and hardware manufacturers. He too found ISTJ (23%) to be the most common type, with INTJ (15%) in second, and INTP (12%) a close third. He also noted that these three types comprised 50% of his sample. Lyons also found thinking (81%) and judging (65%) types to be in the majority; furthermore, 67% of his subjects were introverts. Lyons was the first to observe that R&D companies that do a lot of state-of-the-art development attract and hire more Ns than Ss. The opposite occurs in information systems departments of ordinary companies, where the bulk of the work involves maintaining and enhancing production systems.

Smith (1989) assessed 37 systems analysts (information systems professionals) at a large insurance company in South Africa. The most frequent types in the sample were ISTJ (35%) and ESTJ (30%), there were slightly more introverts (57%), with a heavy bias towards the sensing (81%), thinking (89%), and judging (86%). Interestingly, the four NF combinations were not present at all in this small sample. Larger and diverse samples would allow more comprehensive data and definitive conclusions.

The common thread running through the results of these studies is the prevalence of introversion, thinking, judging, and almost as many sensing as intuitive types among software professionals. In the past, it seemed reasonable to think of computer work as a practical application of mathematical concepts, as in the aerospace industry, but this is no longer true. Today, software permeates almost all activities of modern society, a fact which makes the software engineering a very broad field of study as opposed to the specialized scientific programming of a few decades ago. Software developers can act in occupations without knowing or using mathematics; consequently, the profile of software engineering students might have changed.

In addition, software engineering has become a broad field of study; as a result, some skills necessary to work successfully in this area 30 years ago may no longer apply. For example, software design is much more than manipulating formal or semiformal notations. It has everything to do with interactions between designers and users, i.e., the designer's perception of what the user wants, and the user's perception of what he/she really needs, and vice versa. Nowadays, successful software applications are those developed after a tremendous amount of time has been spent with the user in the form of prototyping, experimenting, and feedback. This is the proper development life cycle of any useful software system. Obviously, with the change in demands on software engineers, further research is needed to establish an up-to-date profile of software developers.

## Method

The sample consisted of 68 Brazilian software engineering students attending universities in the state of Sao Paulo, Brazil. They were all enrolled in upper level computer science or software engineering courses and were administered the MBTI (Form G, in Portuguese language) to determine their personality types. There were invited to take the MBTI either at home or at the university, but not in a class setting. The students were selected to take part in this survey based on their solid background and interest in software development; but their GPAs were not taken into account. There was, however, a disproportionate presence of men (81%) in the sample.

**Results**

The type distribution of the software engineering students is summarized in Table 1. The SRTT has been calculated for that table based on available data about the profile of a national sample of 36,437 subjects of the Brazilian population (Fellipelli, Saad & Vizioli, 2002).

It is noteworthy the two highest self-selection index ratios occurred with ISFP ($I = 3.65$) and INTP ($I = 3.59$) ($p<.001$). Incidentally, ENTJ accounted for 4.4% of the subjects in that sample compared to 14.5% in the general population. Hence, there is a dramatic difference between the ENTJ ($I = 0.30$) percentage in the general population and the same type in our sample.

It can been concluded that the type distribution of software engineering students in Brazil is different from the type distribution found in a general population of that country, as demonstrated by the variation of the self-selection index ratio within the range of 0.30 to 3.65.

**Table 1. Type Distribution of Brazilian Software Engineering Students
and SRTT Comparison to a Sample of the Brazilian Population.**

N = 68,  + = 1% of N,  I = Selection Ratio Index,  *p<.05  **p<.01  ***p<.001

| The Sixteen Complete Types | | | | Dichotomous Preferences | | | |
|---|---|---|---|---|---|---|---|
| ISTJ<br>n = 13<br>(19.12%)<br>I =1.34<br>+ + + + +<br>+ + + + +<br>+ + + + +<br>+ + + + | ISFJ<br>n = 2<br>(2.94%)<br>I = 1.17<br>+ + + | INFJ<br>n = 1<br>(1.47%)<br>I = 1.12<br>+ | INTJ<br>n = 5<br>(7.35%)<br>I = 1.14<br>+ + + + +<br>+ + | E<br>I<br><br>S<br>N<br><br>T<br>F<br><br>J<br>P | 30<br>38<br><br>40<br>28<br><br>54<br>14<br><br>35<br>33 | (44.1%)<br>(55.9%)<br><br>(58.8%)<br>(41.2%)<br><br>(79.4%)<br>(20.6%)<br><br>(51.5%)<br>(48.5%) | ***I=0.68<br>***I=1.60<br><br>I=1.04<br>I=0.95<br><br>I=0.99<br>I=1.05<br><br>**I=0.75<br>**I=1.53 |
| ISTP<br>n = 3<br>(4.41%)<br>I = 1.22<br>+ + + + | ISFP<br>n = 3<br>(4.41%)<br>I = 3.65*<br>+ + + + | INFP<br>n = 2<br>(2.94%)<br>I = 1.65<br>+ + + | INTP<br>n = 9<br>(13.24%)<br>I = 3.59***<br>+ + + + +<br>+ + + + +<br>+ + + | \multicolumn{4}{l}{Pairs and Temperaments} |
| | | | | IJ<br>IP<br>EP<br>EJ<br><br>ST<br>SF<br>NF<br>NT<br><br>SJ<br>SP<br>NP<br>NJ<br><br>TJ<br>TP<br>FP<br>FJ<br><br>IN<br>EN<br>IS<br>ES<br><br>ET<br>EF<br>IF<br>IT | 21<br>17<br>16<br>14<br><br>32<br>8<br>6<br>22<br><br>25<br>15<br>18<br>10<br><br>29<br>25<br>8<br>6<br><br>17<br>11<br>21<br>19<br><br>24<br>6<br>8<br>30 | (30.9%)<br>(25.0%)<br>(23.5%)<br>(20.6%)<br><br>(47.1%)<br>(11.8%)<br>(8.8%)<br>(32.4%)<br><br>(36.8%)<br>(22.1%)<br>(26.5%)<br>(14.7%)<br><br>(42.7%)<br>(36.8%)<br>(11.8%)<br>(8.8%)<br><br>(25.0%)<br>(16.2%)<br>(30.9%)<br>(27.9%)<br><br>(35.3%)<br>(8.8%)<br>(11.8%)<br>(44.1%) | I=1.26<br>***I=2.43<br>I=1.10<br>***I=0.47<br><br>I=0.99<br>I=1.27<br>I=0.85<br>I=0.98<br><br>I=0.85<br>*I=1.61<br>I=1.47<br>*I=0.58<br><br>**I=0.73<br>**I=1.66<br>I=1.24<br>I=0.87<br><br>**I=1.89<br>*I=0.54<br>I=1.43<br>I=0.79<br><br>I=n.a.<br>I=n.a.<br>I=n.a.<br>I=n.a. |
| ESTP<br>n = 8<br>(11.76%)<br>I = 1.76*<br>+ + + + +<br>+ + + + +<br>+ + | ESFP<br>n = 1<br>(1.47%)<br>I = 0.67<br>+ | ENFP<br>n = 2<br>(2.94%)<br>I = 0.68<br>+ + + | ENTP<br>n = 5<br>(7.35%)<br>I = 0.90*<br>+ + + + +<br>+ + | | | | |
| ESTJ<br>n = 8<br>(11.76%)<br>I = 0.51*<br>+ + + + +<br>+ + + + +<br>+ + | ESFJ<br>n = 2<br>(2.94%)<br>I = 0.88<br>+ + + | ENFJ<br>n = 1<br>(1.47%)<br>I = 0.50<br>+ | ENTJ<br>n = 3<br>(4.41%)<br>I = 0.30*<br>+ + + + | | | | |

| Jungian Types (E) | | | Jungian Types (I) | | | Dominant Types | | | . | |
|---|---|---|---|---|---|---|---|---|---|---|
| | n | % | | n | % | | n | % | . | |
| E-TJ | 11 | 16.2% | I-TP | 12 | 17.6% | Dt. T | 23 | 33.8% | . | *Luiz Fernando Capretz* |
| E-FJ | 3 | 4.4% | I-FP | 5 | 7.4% | Dt. F | 8 | 11.9% | . | *Psychological Types of* |
| ES-P | 9 | 13.2% | IS-J | 15 | 22.1% | Dt. S | 24 | 35.3% | . | *Brazilian Software* |
| EN-P | 7 | 10.3% | IN-J | 6 | 8.8% | Dt. N | 13 | 19.1% | . | *Engineering Students.* |